\documentclass[conference]{IEEEtran}
\usepackage{amssymb}
\usepackage{amsfonts}
\usepackage{times,amsmath,epsfig}
\usepackage{color}
\usepackage{ulem}
\usepackage{latexsym}

\begin{document}
%
\title{\huge Knowledge-Aided Reweighted Belief Propagation LDPC Decoding using Regular and Irregular Designs}

\author{\IEEEauthorblockN{Jingjing Liu}
\IEEEauthorblockA{Department of Electronics\\
The University of York\\
Heslington, York, YO10 5DD, UK \\
Email: jl622@ohm.york.ac.uk} \and \IEEEauthorblockN{Rodrigo C. de
Lamare}
\IEEEauthorblockA{Department of Electronics\\
The University of York\\
Heslington, York, YO10 5DD, UK \\
Email: rcdl500@ohm.york.ac.uk}}

\maketitle

\begin{abstract}
In this paper a new message passing algorithm, which takes advantage
of both tree-based re-parameterization and the knowledge of short
cycles, is introduced for the purpose of decoding LDPC codes with
short block lengths. The proposed algorithm is called variable
factor appearance probability belief propagation (VFAP-BP) algorithm
and is suitable for wireless communications applications, where both
good decoding performance and low-latency are expected. Our
simulation results show that the VFAP-BP algorithm outperforms the
standard BP algorithm and requires a significantly smaller number of
iterations than existing algorithms when decoding both regular and
irregular LDPC codes.
\end{abstract}


\section{Introduction}
Low-density parity-check (LDPC) codes were first introduced by
Robert Gallager in his doctoral dissertation \cite {Gallager} and
re-discovered by MacKay, Luby, and others in the 1990s \cite
{MacKay}, \cite{luby}. It has been widely recognized that LDPC codes
are able to closely approach the channel capacity by using iterative
decoding algorithms, which are parallelizable in hardware and have
much lower per-iteration complexity than turbo codes \cite {Moon}.
Unlike turbo codes, it is very easy to implement LDPC codes with any
block length and flexible code rate due to the convenience of
adjusting the size of the parity-check matrix. Moreover, most of
decoding errors are detectable since the decoded codeword is
validated by a simple set of parity-check equations. Equipped with
efficient decoders, LDPC codes have found applications in a number
of communication standards, such as DVB-S2, IEEE 802.16 and Wi-Fi
802.11. Nevertheless, the decoding algorithms of LDPC codes normally
require a significantly higher number of iterations than that of
turbo codes, which results in severe decoding latency \cite {Moon}.

The belief propagation (BP) algorithm is an efficient message
passing algorithm which has been employed to solve a variety of
inference problems in wireless communications, among which its
applications in decoding powerful error-correcting codes are the
most noticeable. Various versions of BP-based algorithms
\cite{berrou}-\cite{xiao} were reported for decoding turbo codes and
LDPC codes. All relevant decoding strategies, either mitigating the
error floor or improving waterfall behavior, can be classified into
two categories: 1) removing the short cycles in the code graph to
avoid ``near-codeword" or ``trapping sets"; 2) enhancing the
suboptimal BP decoding algorithm, when using maximum-likelihood (ML)
decoding is intractable \cite {yedidia2}. However, in wireless
communications where a large amount of data transmission and data
storage are required, those decoding algorithms fail to guarantee
convergence and still suffer from high-latency due to the fact that
too many iterations are often required. In \cite{wainright} and
\cite {wainright2}, the authors state that the BP algorithm is
capable of producing the exact inference solutions when the
graphical model is a spinning tree, while it is not guaranteed to
converge if the graph possesses cycles which significantly
deteriorate the overall performance. Inspired by the tree-reweighted
BP (TRW-BP) algorithm \cite{wainright}, Wymeersch and others
\cite{Wymeersch} recently proposed the uniformly reweighted BP
(URW-BP) algorithm which takes advantage of BP's distributed nature
and defines the factor appearance probability (FAP) in
\cite{wainright} as a constant value. In \cite {Wymeersch2}, the
URW-BP has been shown to consistently outperform the standard BP in
terms of LDPC decoding among other applications.

In this paper, we explore the re-parameterization of a certain part
of a factorized representation of the graphic model while also
statistically taking the effect of short cycles into account. By
combining the re-parameterization framework with the knowledge about
the structure of cycles of a graph obtained by the cycle counting
algorithm \cite {Halford}, which has been successfully employed in
our previous works on rate-compatible LDPC codes \cite {Liu}, \cite
{Liu2}, we present the variable FAP BP (VFAP-BP) algorithm which
aims to decode regular and irregular LDPC codes with short block
lengths more effectively. The main contributions are:

\begin{itemize}
\item A knowledge-aided BP algorithm is devised such that the
reweighting factors (FAPs) are chosen by a simple criterion.
Moreover, the proposed algorithm can be applied to both symmetrical
and asymmetrical graphs.


\item We conduct a study on the most recent reweighted BP algorithms
\cite {Wymeersch}, \cite {Wymeersch2}, and compare the proposed
VFAP-BP algorithm to the standard BP and URW-BP algorithms, in terms
of convergent behavior as well as decoding performance.


\end{itemize}

The organization of this paper is as follows: Section
\uppercase\expandafter{\romannumeral 2} introduces the background of
decoding LDPC codes using standard BP message passing rules and a
tree-based re-parameterization method. In Section
\uppercase\expandafter{\romannumeral 3}, the proposed VFAP-BP
algorithm is presented in detail. Section
\uppercase\expandafter{\romannumeral 4} shows the simulation results
with analysis and discussions. Finally, Section
\uppercase\expandafter{\romannumeral 5} concludes the paper.

\section{Graphical Representation and BP Message Passing Rules}

\begin{figure}[htb]
\begin{minipage}[h]{1.0\linewidth}
  \centering
  \centerline{\epsfig{figure=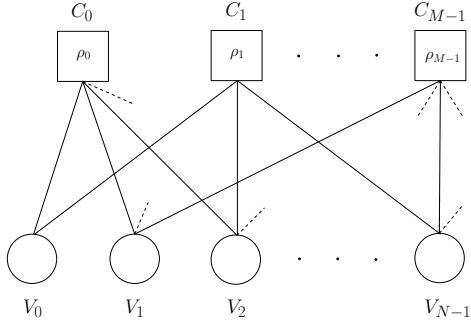,scale=0.3}} \vspace{-0em}\caption{The graphical model depicts BP decoding algorithms for LDPC codes, where $\rho_i (i=0, 1, \ldots, {M-1})=1$ corresponds to the standard BP, $\rho_i (i=0, 1, \ldots, {M-1})=\rho_{u}$ corresponds to the URW-BP, and $\rho_i (i=0, 1, \ldots, {M-1})=\rho_{v}$ or $1$ depending on a variable condition corresponds to the proposed VFAP-BP.} \label{fig:graphmodel}
\end{minipage}
\end{figure}

The advantages of LDPC codes arise from its sparse (low-density)
parity-check matrices which can be uniquely depicted by graphical
representations, referred as Tanner graphs \cite {tanner}. For
instance, an $M \times N$ sparse matrix $\boldsymbol{H}$ can be
represented by a bipartite graph $G$, as in Fig. \ref
{fig:graphmodel} where $C_0, C_1, \ldots, C_{M-1}$ denote parity
check nodes corresponding to $M$ parity check equations and $V_0,
V_1, \ldots, V_{N-1}$ denote variable nodes corresponding to $N$
encoded bits. There is an edge connecting the check node $C_i$ and
the variable node $V_j$ in the factor graph if the entry $h_{ij}$ of
the parity-check matrix $\boldsymbol{H}$ equals $1$. Suppose we have
$K$ information bits being transmitted and a set of codewords
$\boldsymbol {x}$ with block length $N$ is formed under the code
rate $R$ is $K/N$ by an LDPC encoder. At the receiver side, the
decoder strives to find an $1 \times N$ estimated codeword
$\boldsymbol {\hat x}$ which satisfies the parity-check condition
$\boldsymbol{H {\hat x}^T}=\boldsymbol{0}$. Thus, we can interpret
the decoding process as finding $\boldsymbol{\hat x}=\arg \max
p(\boldsymbol {x}|\boldsymbol{y})$. Using Bayes' rule the a
posteriori distribution becomes

\begin{equation} \label{1}
\centering
{p(\boldsymbol{x}|\boldsymbol{y})=\frac{p(\boldsymbol{y}|\boldsymbol{x})p(\boldsymbol{x})}{p(\boldsymbol{y})}},
\end{equation}
where the likelihood ratios $p(\boldsymbol{y}|\boldsymbol{x})$ can
be obtained from the channel and $p(\boldsymbol{x})$ is the prior
information. Nevertheless, directly calculating
$p(\boldsymbol{x}|\boldsymbol{y})$ or $p(\boldsymbol{y})$ is
computationally prohibitive because of the size of $\boldsymbol x$
\cite {wainright2}. For this reason, we resort to BP as a
near-optimal message passing algorithm which can approximate either
$p(\boldsymbol{x}|\boldsymbol{y})$ or $p(\boldsymbol{y})$ \cite
{yedidia}.

\subsection{Standard BP Algorithm for Decoding LDPC Codes}

The BP algorithm is a powerful algorithm to approximately solve
inference problems in decoding LDPC codes. This message passing
algorithm computes accurate marginal distributions of variables
corresponding to each node of a graphical model, and is
exceptionally useful when optimal inference decoding is
computationally prohibitive due to the large size of a graph \cite
{wainright}. Additionally, the BP algorithm is capable of producing
the exact inference solutions if the graphical model is acyclic
(i.e., a tree), but the convergence is no longer guaranteed when the
graph possesses short cycles. The BP algorithm for computing
$p(x_j|\boldsymbol{y})$ for $(j=0, 1, \ldots, {N-1})$ is a
distributed algorithm. As shown in Fig. \ref {fig:graphmodel}, all
the check nodes and the variable nodes work cooperatively and
iteratively so as to estimate $p(x_j|\boldsymbol{y})$ for $(j=0, 1,
\ldots, {N-1})$ \cite {ryan and lin}. In addition, for achieving a
numerical stability as well as saving storage space the marginal
distribution $p(x_j|\boldsymbol{y})$ is replaced by log-likelihood
ratios (LLRs) $L(x_j|\boldsymbol{y})\triangleq
\log{\frac{p(x_j=1|\boldsymbol{y})}{p(x_j=0|\boldsymbol{y})}}$.

Consequently, we derive the following message passing rules of
standard BP over an additive white Gaussian noise (AWGN) channel:
all variable nodes $V_j$ for $(j=0, 1, \ldots, {N-1})$ are
initialized by
\begin{equation} \label{2}
\centering
{\Psi_{ji}=L(x_j)=\log{\frac{p(y_j|x_j=1)}{p(y_j|x_j=0)}}=2\frac{y_j}{\sigma^{2}}},
\end{equation}
where $\sigma^{2}$ is the noise variance. Then for all check nodes
$C_i$ for $(i=0, 1, \ldots, {M-1})$, we update the message sent from
$C_i$ to $V_j$ as:
\begin{equation} \label{3}
\centering {\Lambda_{ij}=2 \mathrm {tanh}^{-1}\big{(}\prod_{j'\in
\mathcal{N}(i)\backslash j}\mathrm{tanh}
\frac{\Psi_{j'i}}{2}}\big{)},
\end{equation}
where `$\rm {tanh}(\cdot)$` denotes the hyperbolic tangent function
and $\mathcal{N}(i)\backslash j$ denotes the neighboring variable
nodes' set of $C_i$ except $V_j$. Next, we update the message sent
from $V_j$ to $C_i$ for all variable nodes $V_j$ by:
\begin{equation} \label{4}
\centering {\Psi_{ji}=L(x_j)+\sum_{i'\in \mathcal{N}(j)\backslash
i}\Lambda_{i'j}},
\end{equation}
where $i'\in \mathcal{N}(j)\backslash i$ is the neighboring set of
check nodes of $V_j$ except $C_i$. Finally, at every variable node
$V_j$ we acquire the so-called beliefs
\begin{equation} \label{5}
\centering {b(x_j)=L(x_j)+\sum_{i\in \mathcal{N}(j)}\Lambda_{ij}},
\end{equation}
which is exactly the approximation of $L(x_j|\boldsymbol{y})$ and
\begin{equation}\label{6}
\hat{x}_j=\left\lbrace \begin{array}[c]{c}\begin{split}
& 1, \;{\rm if} \;{\rm and}\; {\rm only}\; {\rm if}\; b(x_j)>0\\
& 0, \;{\rm if} \;{\rm and}\; {\rm only}\; {\rm if}\; b(x_j)<0
\end{split}
\end{array}
\right.
\end{equation}


While applying the above message passing rules, the variable nodes
(check nodes) process the incoming message and send the extrinsic
information to their neighboring check nodes (variable nodes) back
and forth in an iterative fashion, until all $M$ parity check
equations are satified ($\boldsymbol{H {\hat x}^T}=\boldsymbol{0}$),
or the decoder reaches the maximum number of iterations.

\subsection{Tree-Based Re-parameterization and Bethe's Entropy}
When the factor graph contains short cycles, the standard BP
algorithm normally requires a larger number of iterations but still
fails to converge \cite{MacKay}, \cite{yedidia}, \cite{wainright}.
To address that problem, Wainright \textit {et al.} developed the
TRW-BP algorithm which improves the convergence of BP by reweighting
certain portions of the factorized graphical representation in
\cite{wainright} and \cite{wainright2}. However, TRW-BP algorithm
only considers a factorized graph with pairwise interactions, i.e.,
Markov field, and is not suitable for the distributed inference
problem. The URW-BP algorithm \cite {Wymeersch}, \cite {Wymeersch2}
extends the pairwise factorizations of TRW-BP to hypergraphs and
reduces a series of globally optimized parameters to a simple
constant.

Given a factor graph, the Kullback Leibler divergence \cite{cover}
between the beliefs $b(\boldsymbol x)$ in (\ref{5}) and
$p(\boldsymbol{x}|\boldsymbol{y})$ is defined as
\begin{equation} \label{7}
\centering {{\rm KL}(b||p)=\sum_{\boldsymbol {x}}b(\boldsymbol
x)\log\frac{b(\boldsymbol x)}{p(\boldsymbol{x}|\boldsymbol{y})}\geq
0}.
\end{equation}
Combining the above equation with (\ref{1}), we have the inequality
\begin{equation} \label{8}
\centering {\log p(\boldsymbol y)\geq
\mathcal{H}(b)+\mathcal{F}(b)},
\end{equation}
in which $\mathcal{H}(b)$ is the entropy of the distribution
$b(\boldsymbol x)$ and $\mathcal{F}(\cdot)$ is the factorization
function. In addition, (\ref {8}) is valid with equality if and only
if $b(\boldsymbol x)=p(\boldsymbol{x}|\boldsymbol{y})$ (more details
can be found in \cite{Wymeersch2}). Since the fixed points of the BP
algorithm correspond to the stationary points of Bethe's free energy
\cite {yedidia}, the entropy term in (\ref {8}) can be replaced by
Bethe's approximation with a constant reweighting factor $\rho_{u}$
as
\begin{equation} \label{9}
\centering
{\mathcal{H}(b|\rho_u)=\sum_{j=1}^{N}\mathcal{H}(b_j)-\sum_{i=1}^{M}\rho_u\mathcal{I}_{\mathcal{N}(i)}(b_{\mathcal{N}(i)})}.
\end{equation}
where $\mathcal{I}_{\mathcal{N}(i)}(b_{\mathcal{N}(i)})$ is the
mutual information term. The work in \cite {Wymeersch2} points out
that maximizing Bethe's entropy is equivalent to maximizing the
entropies of $b(\boldsymbol x)$ as well as minimizing the dependence
among all variables. Thus, a new message passing rule can be
obtained by finding stationary points of the Lagrangian in (\ref
{8}). It is also important to note that $\rho_u = 1$ corresponds to
the standard BP algorithm.

\section{Proposed VFAP-BP Decoding Algorithm }
\begin{table}[!t]
\centering
    \caption{The algorithm flow of the VFAP-BP algorithm}
    \label {tab:VFAP-BP}     
    \begin{small}
        \begin{tabular}{l l}

\hline \\

\bf{Initialization:} \\ \\
1: Run (\ref{10})-(\ref{15}), using (\ref{16}) to find the girth $g$
and $s_i$ the number \\of length-$g$ cycles passing the check node $C_i$ ; \\

2: Determine variable FAPs for each check node: if $s_i<\mu_g$
$\rho_i=1$, \\otherwise $\rho_i=\rho_v$ where
$\rho_v=2/\bar{n_D}$;\\ \\

\bf{VFAP-BP decoding:}\\ \\
Step 1: Set $I_{max}$ the maximum number of iterations and
initialize \\$L(\boldsymbol x)=2\frac{\boldsymbol y}{\sigma^2}$; \\
Step 2: Update the message passed from variable node $V_j$ to
check\\
node $C_i$ using (\ref{17}), where $\Lambda_{i'j}$ and
$\Lambda_{ij}$ are $0s$ at the first iteration;\\
Step 3: Update the message passed from variable node $C_i$ to
check \\node $V_j$ using (\ref{3});\\
Step 4: Update the belief $b(x_j)$ using (\ref{18}) and decide $\boldsymbol {\hat{x}}$; \\
Step 5: Decoding stops if $\boldsymbol{H {\hat x}^T}=\boldsymbol{0}$
or
$I_{max}$ is reached, otherwise \\go back to Step 2.\\

\hline \\
    \end{tabular}
    \end{small}
\end{table}

In this section, the proposed VFAP-BP algorithm is presented which
selects the reweighting parameters under a simple criterion. As
mentioned before, the proposed VFAP-BP algorithm does not require a
symmetrical factor graph and, for this reason, it is eligible for
LDPC codes with both regular and irregular designs. In the
following, we briefly explain the cycle counting algorithm then
introduce our message passing rules and the VFAP-BP decoding
algorithm flow.


The cycle counting algorithm \cite {Halford} transforms the problem
of counting cycles into that of counting the so-called lollipop
walks, and has been used in our previous works on rate-compatible
LDPC codes \cite {Liu}, \cite {Liu2}. Given a bipartite graph $G(V,
E)$ where $V$ denotes the set of vertices ($V= V_{c} \bigcup
V_{s}$), $E$ denotes the set of edges, and `$\mid\cdot\mid$`
represents the cardinality of a set. Define $\boldsymbol
{P}_{2k}^{v_{c}}$ as a $\mid{{V_{c}}}\mid \times \mid{{V_{c}}}\mid$
matrix in which the $(i,j)th$ element is the number of paths of
length $2k$ from $v_{c_{i}}\in {V_{c}}$ to $v_{c_{j}}\in {V_{c}}$.
Similarly, define ${\boldsymbol{P}_{2k+1}^{v_{c}}}$ as a
$\mid{{V_{c}}}\mid \times \mid{{V_{s}}}\mid$ matrix in which the
$(i,j)th$ element is the number of paths of length $2k+1$ from
$v_{c_{i}}\in {V_{c}}$ to $v_{s_{j}}\in {V_{s}}$. Let us also define
${\boldsymbol{L}_{2k',2k-2k'}^{v_{c}}}$ as a $\mid{{V_{c}}}\mid
\times \mid{{V_{c}}}\mid$ matrix in which the $(i,j)th$ element is
the number of $(2k',2k-2k')$-lollipop walks from $v_{c_{i}}\in
{V_{c}}$ to $v_{c_{j}}\in {V_{c}}$. Similarly, define
${\boldsymbol{L}_{2k'+1,2k-2k'}^{v_{c}}}$ as a $\mid{{V_{c}}}\mid
\times \mid{{V_{s}}}\mid$ matrix in which the $(i,j)th$ element is
the number of $(2k'+1,2k-2k')$-lollipop walks from $v_{c_{i}}\in
{V_{c}}$ to $v_{s_{j}}\in {V_{s}}$. For counting cycles of length
$2k$, the above quantities can be computed by
\begin{equation} \label{10}
\centering {\boldsymbol P_{2k+1}^{v_{c}}=\boldsymbol
P_{2k}^{v_{c}}\boldsymbol E-\sum_{i=0}^{k-1}\boldsymbol
L_{(2i+1,2k-2i)}^{v_{c}}},
\end{equation}
\begin{equation} \label{11}
\centering {\boldsymbol P_{2k}^{v_{c}}=\boldsymbol
P_{2k-1}^{v_{c}}\boldsymbol E^{T}-\sum_{i=0}^{k-1}\boldsymbol
L_{(2i,2k-2i)}^{v_{c}}},
\end{equation}
\begin{equation} \label{12}
\centering {\boldsymbol P_{2k+1}^{v_{s}}=\boldsymbol
P_{2k-1}^{v_{s}}\boldsymbol E^{T}-\sum_{i=0}^{k-1}\boldsymbol
L_{(2i+1,2k-2i)}^{v_{s}}},
\end{equation}
\begin{equation} \label{13}
\centering {\boldsymbol P_{2k}^{v_{s}}=\boldsymbol
P_{2k-1}^{v_{s}}\boldsymbol E-\sum_{i=0}^{k-1}\boldsymbol
L_{(2i,2k-2i)}^{v_{s}}},
\end{equation}
\begin{equation} \label{14}
\centering {\boldsymbol L_{(0,2k)}^{v_{c}}=(\boldsymbol
P_{2k-1}^{v_{c}}\boldsymbol E^{T})\circ \boldsymbol I},
\end{equation}
\begin{equation} \label{15}
\centering {\boldsymbol L_{(0,2k)}^{v_{s}}=(\boldsymbol
P_{2k-1}^{v_{s}}\boldsymbol E)\circ \boldsymbol I},
\end{equation}
where `$\circ$` means element-wise matrix multiplication,
$\boldsymbol E$ is the edge matrix and $\boldsymbol I$ is the
identity matrix. The total number of cycles of length $2k$ is
\begin{equation} \label{16}
\centering { N_{2k}=\frac{1}{2k}\mathrm {Tr}\big(\boldsymbol
L_{(0,2k)}^{v_{c}}\big)=\frac{1}{2k}\mathrm {Tr}\big(\boldsymbol
L_{(0,2k)}^{v_{s}}\big)},
\end{equation}
where `$\rm {Tr}(\cdot)$` means the trace of a related matrix. In
order to find the girth $g$ and to count cycles of length $g$, $g+2$
and $g+4$ in a Tanner Graph, \eqref{10}-\eqref{15} are expanded and
updated with each other such that counting short cycles is
equivalent to counting the so-called lollipop recursions \cite
{Halford}.

In this work, we only focus on the value of $g$, $s_i$ for $i=0, 1,
\ldots, {M-1}$ is the number of length-$g$ cycles passing a check
node $C_i$, and $\mu_g$ is the average number of length-$g$ cycles
passing a check node. In a similar way to \cite {wainright2} and
\cite{Wymeersch2}, the reweighting vector $\rho_i=[\rho_0, \rho_1,
\ldots, \rho_{M-1}]$ consists of variable factor appearance
probabilities (FAP), which describe the probabilities of any check
node appearing in a potential spinning tree. As shown in Fig.
\ref{fig:graphmodel}, every check node $C_i$ is assigned to a FAP
value such that the message from each check node is partially
reweighted. Note that when $\rho_i (i=0, 1, \ldots, {M-1})=1$ it is
equivalent to the standard BP, and when $\rho_i (i=0, 1, \ldots,
{M-1})=\rho_{u}$ it is equivalent to the URW-BP. The message passing
rules of the proposed VFAP-BP algorithm can be described as follows.
Firstly, the message sent from $V_j$ to $C_i$ is given by
\begin{equation} \label{17}
\centering {\Psi_{ji}=L(x_j)+\rho_{i'}\sum_{i'\in
\mathcal{N}(j)\backslash i}\Lambda_{i'j}-(1-\rho_i)\Lambda_{ij}}.
\end{equation}
The message sent from $C_i$ to $V_j$ is the same as in (\ref{3}),
and we have the belief $b(x_j)$ with respect to $x_j$ described by
\begin{equation} \label{18}
\centering {b(x_j)=L(x_j)+\rho_i\sum_{i\in
\mathcal{N}(j)}\Lambda_{ij}}.
\end{equation}

Using the above message passing rules, the proposed VFAP-BP decoding
algorithm is depicted in Table \ref {tab:VFAP-BP}. Note that $\rho_v
= 2/ n_D$ at the initialization is an estimate of the optimized FAP
value according to \cite{wainright2}, where $n_D$ is the average
connectivity for $N$ variable nodes. when compared to existing
re-parameterization techniques \cite{wainright}, \cite{Wymeersch},
the proposed VFAP-BP algorithm only needs an optimization of short
cycles with complexity ($\mathcal{O}(gN)$) instead of a global
optimization with complexity ($\mathcal{O}(M^{2}N)$).



\section{Simulation Results}

In this section, we compare the proposed VFAP-BP algorithm with the
URW-BP algorithm \cite {Wymeersch} and the standard BP algorithm
while decoding LDPC codes with a short block length. The LDPC codes
are designed by the PEG \cite{Hu} method, having a block length of
$500 (N=500)$ and a code rate of $1/2$. Other designs and
improvements over the PEG are also possible \cite{ace_peg,dopeg}.
The average connectivity of $N$ variable nodes is derived as
\begin{equation} \label{19}
\centering {
n_d=\frac{1}{\int_{0}^{1}{\lambda(x)}dx}=\frac{M}{N\int_{0}^{1}{\nu(x)}dx}},
\end{equation}
in which $\lambda(x)$ denotes the degree distribution of variable
nodes and $\nu(x)$ denotes the degree distribution of check nodes.
For the regular code tested, the degree of variable codes is $4
(\lambda(x)=x^3)$, the degree of check nodes is $6 (\nu(x)=x^5)$,
and the average connectivity $n_{d,reg}$ is $4$. For the irregular
code, the degree distribution of variable nodes is
$\lambda(x)=0.21\times x^5+0.25\times x^3+0.25\times x^2+0.29\times
x$, the degree distribution of check nodes is $\nu(x)=x^5$, and the
average connectivity $n_{d,irreg}$ is $3$. By using the counting
cycle algorithm we found that there are $964$ length-$6$ cycles in
the regular graph while there are $1260$ length-$8$ cycles in the
irregular graph. As described in Section
\uppercase\expandafter{\romannumeral 3}, $\rho_i \in \boldsymbol
{\rho}=[\rho_0, \rho_1, \ldots, \rho_{M-1}]$ equals $1$ if
$s_i<\mu_g$, and equals $2/n_d$ if $s_i>\mu_g$.

\begin{figure}[!htb]
\begin{center}
\def\epsfsize#1#2{1.0\columnwidth}
\epsfbox{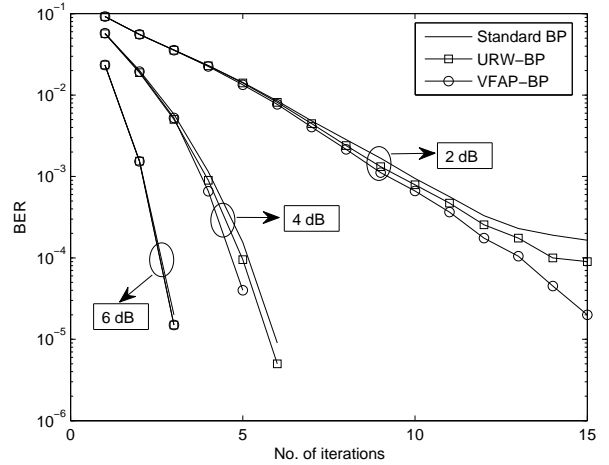} \vspace{-1.5em}\caption{Comparison of the
convergent behaviors of the URW-BP, VFAP-BP and standard BP
algorithms for decoding regular LDPC codes, where SNR equal to $2$
dB, $4$ dB and $6$ dB.}\label{fig:regconver}
\end{center}
\end{figure}

\begin{figure}[!htb]
\begin{center}
\def\epsfsize#1#2{1.0\columnwidth}
\epsfbox{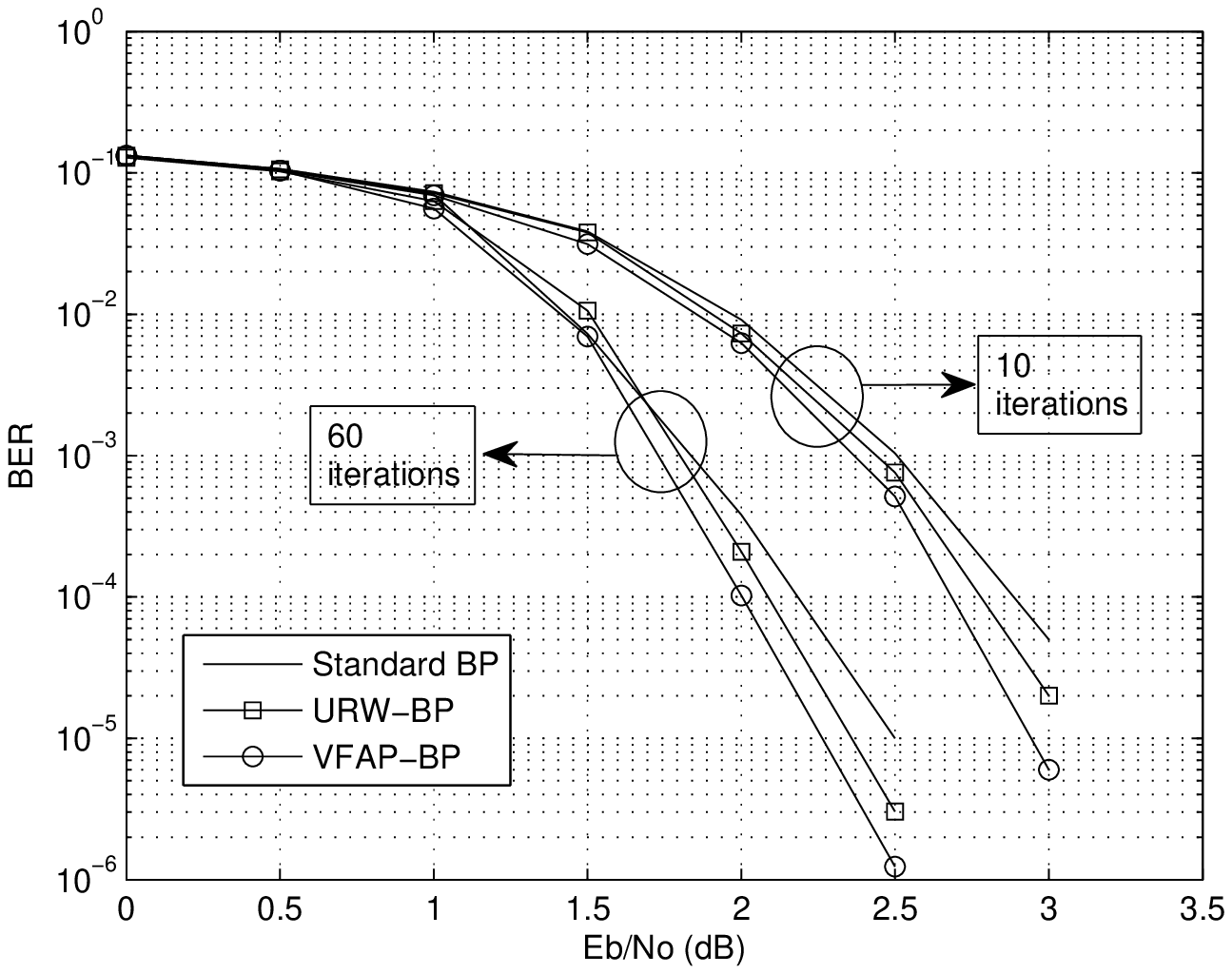} \vspace{-1.5em}  \caption{Comparison of the BER
performances of the VFAP-BP, URW-BP and standard BP algorithms while
decoding regular LDPC codes with $10$ and $60$ maximum decoding
iterations.}\label{fig:regBER}
\end{center}
\end{figure}

\begin{figure}[!htb]
\begin{center}
\def\epsfsize#1#2{1.0\columnwidth}
\epsfbox{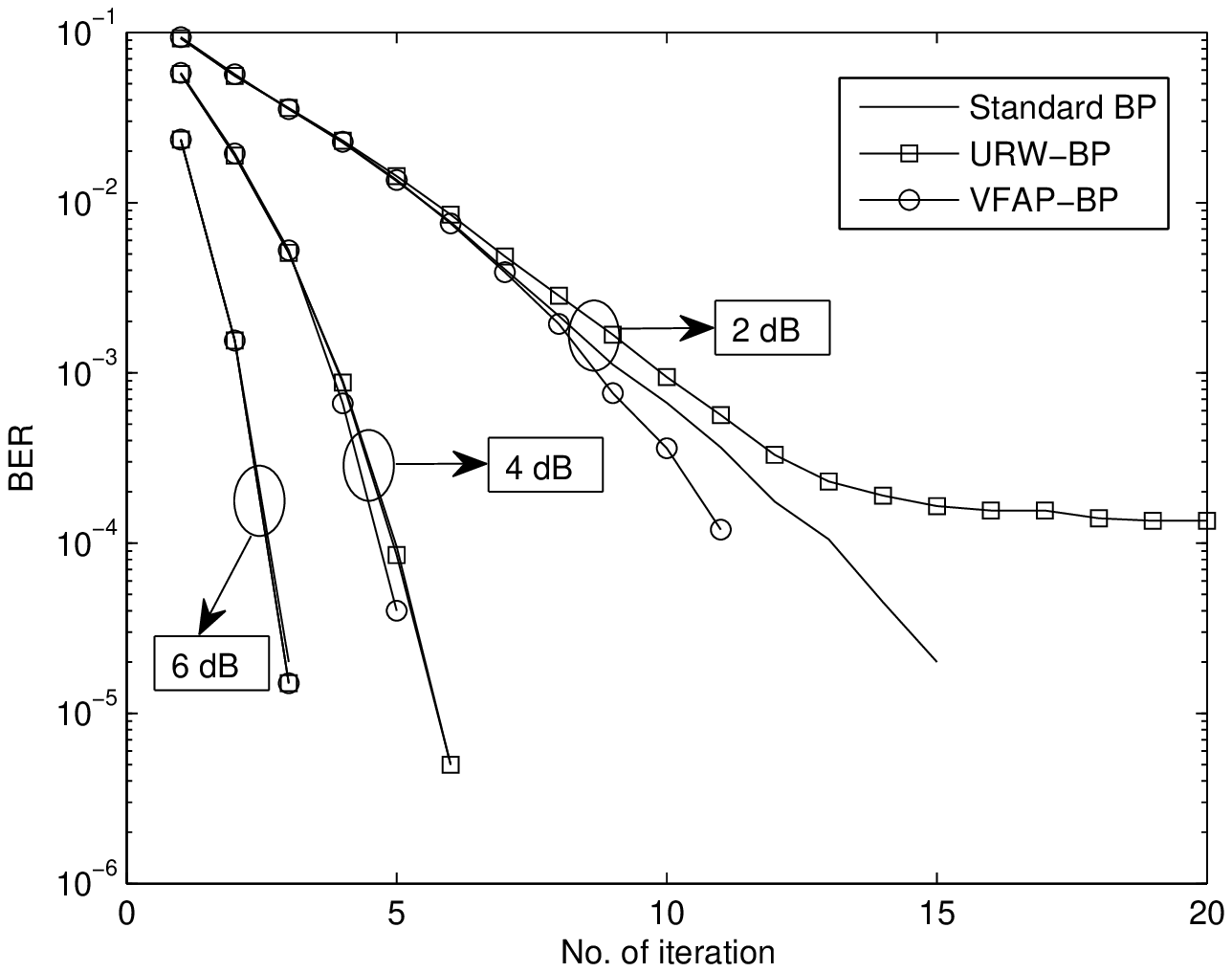} \vspace{-1.5em}\caption{Comparison of the
convergent behaviors of the URW-BP, VFAP-BP and standard BP
algorithms for decoding irregular LDPC codes, where SNR equal to $2$
dB, $4$ dB and $6$ dB.}\label{fig:irregconver}
\end{center}
\end{figure}

\begin{figure}[!htb]
\begin{center}
\def\epsfsize#1#2{1.0\columnwidth}
\epsfbox{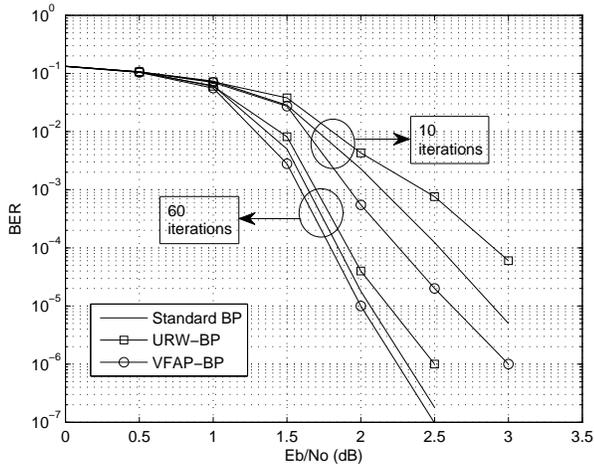} \vspace{-1.5em}  \caption{Comparison of the BER
performances of the VFAP-BP, URW-BP and standard BP algorithms while
decoding irregular LDPC codes with $10$ and $60$ maximum decoding
iterations.}\label{fig:irregBER}
\end{center}
\end{figure}

In Fig. \ref {fig:regconver} the convergent behaviors of the URW-BP,
VFAP-BP and standard BP algorithms are compared, in order to
illustrate that the proposed algorithm converges faster particularly
at lower SNR region. Furthermore, Fig. \ref {fig:regBER} reveals the
decoding performances of three algorithms in which the VFAP-BP
outperforms others regardless of the number of maximum decoding
iterations. When decoding irregular codes with asymmetrical graphs,
as shown in Fig. \ref {fig:irregconver} and in Fig.
\ref{fig:irregBER}, the proposed VFAP-BP algorithm still shows a
better convergence behavior and consistently outperforms the
standard BP, but the URW-BP fails to converge at $2$ dB as well as
no longer outperforms the standard BP with the maximum number
iterations equal to $10$ and $60$, respectively.

\section{Conclusion}

In this paper, we have proposed a message passing decoding algorithm
by exploring the tree-based re-parameterization method and the
knowledge of the presence of short cycles in the graph structure.
The proposed VFAP-BP algorithm has been evaluated when decoding both
short-length regular and irregular LDPC codes. Simulation results
have shown that the proposed VFAP-BP algorithm is capable of
providing good performance while requiring less decoding iterations
than the existing algorithms.


\section*{Acknowledgment}

The authors would like to thank Henk Wymeersch who provided MATLAB
codes of URW-BP algorithm for our comparison purpose.


\begin{thebibliography}{100}
{\footnotesize
\bibitem{Gallager}
R. G. Gallager, ``Low-density parity check codes," \textit {IRE
Trans. Inf. Theory.}, vol. 39, no. 1, pp. 37-45, Jan. 1962.

\bibitem{MacKay}
D. J. C. Mackay and R. M. Neal, ``Near Shannon limit performance of
low density parity check codes," \textit {Electron. Lett.}, vol. 33,
no. 6, pp. 457-458, Mar. 1997.

\bibitem{luby}
N. Alon and M. Luby, ``A linear time erasure-resilient code with
nearly optimal recovery," \textit {IEEE Trans. Information Theory},
vol. 42, no. 11, pp. 1732-1736, Nov. 1996.

\bibitem{Moon}
T. K. Moon, ``Error Correction Coding: Mathematical Methods and
Algorithms," Wiley-Blackwell, July 1, 2005.

\bibitem{berrou}
C. Berrou, A. Glavieux, and P. Thitimajshima, ``Near Shannon limit
error-correcting coding and decoding: Turbo-codes," \textit
{Proceedings IEEE International Conference on Communications,}
Geneva, Switzerland, pp. 1064-1070, May 1993.

\bibitem{richardson}
T. J. Richardson, M. A. Shokrollahi, and R. L. Urbanke, ``Design of
capacity-approaching irregular low-density parity-check codes,"
\textit {IEEE Trans. Information Theory}, vol. 47, no. 2, pp.
619–637, Feb. 2001.

\bibitem{frey}
F.R. Kschischang, B.J. Frey, and H.-A. Loeliger, ``Factor graphs and
the sum-product algorithm," \textit {IEEE Trans. Information
Theory}, vol. 47, no. 2, pp. 498 - 519, Feb. 2001.

\bibitem{xiao}
H. Xiao and A.H. Banihashemi, ``Graph-based message-passing
schedules for decoding LDPC codes," \textit {IEEE Trans.
Communications}, vol. 12, no. 12, pp. 498 - 519, Dec. 2004.

\bibitem{yedidia2}
J. S. Yedidia, W. T. Freeman, and Y. Weiss, ``Understanding belief
propagation and its generalizations," Mitsubishi Electric Res. Labs,
Cambridge, MA, Tech. Rep. TR2001-22, 2002.

\bibitem{wainright}
M. J. Wainwright, T. S. Jaakkola, and A.S. Willsky, ``Tree-based
reparameterization framework for analysis of sum-product and related
algorithms," \textit {IEEE Trans. Information Theory}, vol. 49, no.
5, pp. 1120 - 1146, May 2003.

\bibitem{wainright2}
M. J. Wainwright, T. S. Jaakkola, and A.S. Willsky, ``A new class of
upper bounds on the log partition function," \textit {IEEE Trans.
Information Theory}, vol. 51, no. 7, pp. 2313 - 2335, July 2005.

\bibitem{Halford}
T. R. Halford, K. M. Chugg, ``An algorithm for counting short cycles
in bipartite graph," \textit {IEEE Trans. on Infor. Theory}, vol.
52, no. 1, pp. 287-292, Jan. 2006.

\bibitem{Liu}
J. Liu and R. C. de Lamare, ``Novel Intentional Puncturing Schemes
For Finite-Length Irregular LDPC Codes," \textit {International
Conference on DSP.}, Corfu, Greece, Jul. 2011.

\bibitem{Liu2}
J. Liu and R. C. de Lamare, ``Finite-length rate-compatible LDPC
codes based on extension techniques," \textit {8th International
Symposium on Wireless Communication Systems}, Aachen, Germany, Nov.
2011.

\bibitem{Wymeersch}
H. Wymeersch, F. Penna and V. Savic, ``Uniformly reweighted belief
propagation: A factor graph approach," Information Theory
Proceedings (ISIT), 2011 IEEE International Symposium on Issue Date:
July 31 2011-Aug. 5 2011 On page(s): 2000 - 2004.

\bibitem{Wymeersch2}
H. Wymeersch, F. Penna and V. Savic, ``Uniformly Reweighted Belief
Propagation for Estimation and Detection in Wireless Networks,"
\textit {IEEE Trans. Wireless Communications}, vol. PP, No. 99, pp.
1-9, Feb. 2012.

\bibitem{vfap} J. Liu, R. C. de Lamare, ``Low-Latency
Reweighted Belief Propagation Decoding for LDPC Codes", \textit{IEEE
Communications Letters}, vol. 16, no. 10, pp. 1660-1663, October
2012.

\bibitem{tanner}
R. M. Tanner, ``A recursive approach to low complexity codes,"
\textit {IEEE Trans. Information Theory}, vol. 27, no. 9, pp.
533-547, Sep. 1981.

\bibitem{yedidia}
J. S. Yedidia, W. T. Freeman, and Y. Weiss, ``Understanding belief
propagation and its generalizations," Mitsubishi Electric Res. Labs,
Cambridge, MA, Tech. Rep. TR2001-22, 2002.

\bibitem{ryan and lin}
W. Ryan and S. Lin,``Channel Codes: Classical and Modern," Cambridge
University Press, 1st edition, Oct. 30, 2009.







%
%
%
%
%
%
%
%
%
%
%

\bibitem{cover}
T. M. Cover and J. A. Thomas, ``Elements of Information Theory," 2nd
edition. Wiley-Interscience, 2006.

\bibitem{Hu}
Y. Hu, E. Eleftheriou and D. M. Arnold, "Regular and irregular
progressive edge-growth tanner graph," \textit {IEEE Transaction on
Information Theory}, vol. 51, no. 1, pp. 386-398, Jan. 2005.

\bibitem{ace_peg}
D. Vukobratovic and V. Senk, "Generalized ACE constrained
progressive edge-growth LDPC code design," \textit{IEEE Commun.
Lett.}, vol. 12, no. 1, pp. 32–34, Jan. 2008.

\bibitem{iswcs.11}
A.~G.~D. Uch\^oa, C.~Healy, R.~C.
de~Lamare, and R.~D. Souza, ``LDPC codes based on progressive edge
growth techniques for block fading channels,'' in \emph{Proc. IEEE
{ISWCS}'11}, Aachen, Germany, Nov 2011, pp. 1--5.

\bibitem{uchoa} A. Uchoa, C. Healy, R. C. de Lamare
and R. D. Souza, ``Design of LDPC Codes Based on Progressive Edge
Growth Techniques for Block Fading Channels", \textit{ IEEE
Communications Letters} , vol. 15, no. 11, pp. 1221-1223, 2011.

\bibitem{iswcs.12} A.~G.~D. Uch\^oa, C.~Healy, R.~C.
de~Lamare, and R.~D. Souza, ``Generalised {Quasi-Cyclic} {LDPC}
codes based on progressive edge growth techniques for block fading
channels,'' in \emph{Proc. {IEEE} {ISWCS}'12}, Paris, France, Aug
2012, pp. 1--5.

\bibitem{dopeg} C. T. Healy and R. C. de Lamare,
``Decoder-optimised progressive edge growth algorithms for the
design of LDPC codes with low error floors", \textit{IEEE
Communications Letters}, vol. 16, no. 6, June 2012, pp. 889-892.

}
\end{thebibliography}
\end{document}